\documentclass[showpacs,twocolumn,prl,longbibliography]{revtex4-1}

\usepackage{epsf}
\usepackage{amsmath}
\usepackage{amsfonts}
\usepackage{amssymb}
\usepackage{bm}
\usepackage{bbm}
\usepackage{nicefrac}
\usepackage{color}
\usepackage{maybemath}
\usepackage{pifont}
\usepackage{comment}

\usepackage{graphicx}
\usepackage{epsfig}
\usepackage{latexsym}
\usepackage{pifont}

\def\calL{{\mathcal{L}}}
\def\calP{{\mathcal{P}}}
\def\calQ{{\mathcal{Q}}}

\def\calW{{\mathcal{W}}}

\def\dd{{\mathrm{d}}}
\def\ii{{\mathrm{i}}}
\def\ee{{\mathrm{e}}}

\def\tfrac#1#2{  {\textstyle{#1  \over #2}}   }

\newcommand{\stdrule}{\rule[-2mm]{0mm}{6mm}}

\def\Vdw{Van der Waals}
\def\vdw{van der Waals}

\begin{document}

\title{Virtual Resonant Emission and Oscillatory Long--Range Tails \\
in \vdw{} Interactions of Excited States: QED Treatment and Applications}

\newcommand{\addrROLLA}{Department of Physics,
Missouri University of Science and Technology,
Rolla, Missouri 65409, USA}

\author{U. D. Jentschura}
\author{C. M. Adhikari}
\author{V. Debierre}
\affiliation{\addrROLLA}

\begin{abstract}
We report on a quantum electrodynamic (QED) investigation
of the interaction between a ground state atom
with another atom in an excited state.
General expressions, applicable to any atom, are indicated for the 
long-range tails which are due to
virtual resonant emission and absorption
into and from vacuum modes whose frequency equals 
the transition frequency to available lower-lying atomic states.
For identical atoms, one of which is in an excited state,
we also discuss the mixing term which depends on the symmetry
of the two-atom wave function
(these evolve into either the gerade or the 
ungerade state for close approach),
and we include all nonresonant states in our rigorous QED treatment. 
In order to illustrate the findings, we analyze the
fine-structure resolved \vdw{} interaction 
for $nD$--$1S$ hydrogen interactions with 
$n=8,\,10,\,12$ and find surprisingly large numerical coefficients.
\end{abstract}


\maketitle

{\em Introduction.---}While the long-range interaction between two
ground state atoms has been fully understood
in all interatomic separation regimes
since the work of Casimir and Polder~\cite{CaPo1948},
a completely new situation arises when one of the atoms is
in an excited state~\cite{Ch1972,DeYo1973,AdEtAl2017vdWi,JeEtAl2017vdWii,%
PoTh1995,SaBuWeDu2006}.
In particular, several recent 
studies~\cite{SaKa2015,Be2015,MiRa2015,DoGuLa2015,Do2016}
have reported on long-range,
spacewise-oscillating tails, which decay as slowly as $R^{-2}$
($R$ is the interatomic separation). For excited reference states,
these tails parametrically dominate over the usual
Casimir-Polder type $R^{-7}$ interaction~\cite{CaPo1948}.
Conflicting results have been obtained 
for the oscillating tails~\cite{GoMLPo1966,PoTh1995,SaBuWeDu2006}.
One important question concerns the ratio of 
the oscillatory, resonant terms to the 
non-oscillatory, nonresonant contributions to
\vdw{} interactions,
and the matching and interpolation of the results
with the familiar close-range, nonretarded \vdw{} limit
of the interatomic interaction.
Our aim here is to advance the theory 
of excited-state interatomic interactions,
by including the nonresonant states, 
the dynamically induced correction to the 
atomic decay width (distance-dependent imaginary part 
of the energy shift), and the additional terms
that occur for identical atoms
(namely, the gerade-ungerade mixing term~\cite{Ch1972}).

As an example application, we study a system where a 
highly excited $D$ state interacts with 
a ground state ($1S$) hydrogen atom
(see Refs.~\cite{BiGaJuAl1989,BeEtAl1997,ScEtAl1999}).
In this system, the availability of low-energy $P$ and $F$ states
for virtual dipole transitions from the $nD$ state 
makes the oscillating long-range tails relevant.
An improved understanding is necessary 
for the interpretation of experiments
involving general Rydberg states~\cite{dV2002},
in regard to the determination of fundamental constants.
We concentrate on $nD$--$1S$ interactions with $n=8,10,12$.
The projection- and symmetry-averaged $C_6$
van der Waals coefficient of
the $12D$--$1S$ system amounts to a surprisingly large
numerical value $\langle C_6(12D;1S) \rangle =
227\,756$ in atomic units. SI mksA units are used throughout
this Letter.

{\em Formalism.---}The general idea behind the 
matching of the scattering amplitude and the effective Hamiltonian
has been described in Chap.~85 of Ref.~\cite{BeLiPi1982vol4},
in the context of the interatomic interaction.
In short, one uses the relation
\begin{equation}
\label{match}
\langle \psi'_A, \psi'_B | U( \vec r_A, \vec r_B, \vec R )
| \psi_A, \psi_B \rangle =
\frac{\ii \hbar}{T} \, S_{A'B'AB} \,,
\end{equation}
where $| \psi_A, \psi_B \rangle$ is the initial state of the two-atom system, 
$| \psi'_A, \psi'_B \rangle$ is the final state, and
$H_{\rm eff} = U( \vec r_A, \vec r_B, \vec R )$ is the effective potential
which depends on the electron coordinates $\vec r_i$ (where $i= A,B$ 
denotes the atom). The interatomic distance vector is $\vec R$.
Finally, $T$ is the long time interval which 
results from the integration over the interaction in the 
$S$ matrix formalism [see Eq.~(85.2) of Ref.~\cite{BeLiPi1982vol4}].

It becomes necessary to generalize the 
treatment outlined in Eqs.~(85.1) to~(85.14) 
of Ref.~\cite{BeLiPi1982vol4}
to the case of identical atoms, one of which is 
in an excited state. In this case, 
one has to treat a mixing term~\cite{Ch1972}, which describes 
a scattering process in which the state $| \psi_A, \psi_B \rangle$ 
is scattered into the state $| \psi_B, \psi_A \rangle$;
the two atoms in this case ``interchange'' their quantum states.
The eigenstates of the \vdw{} Hamiltonian~\cite{Ch1972} are states of the form
$(1/\sqrt{2}) \, \left(
| \psi_A, \psi_B \rangle \pm
| \psi_B, \psi_A \rangle \right)$,
and the interaction energy $\Delta E$ is the 
sum of a direct term (which is contained in the 
canonical derivations, e.g., Ref.~\cite{BeLiPi1982vol4}), 
and a mixing term, which is added here and 
whose sign depends on the symmetry of the two-atom state ($\pm$).
We find the following general expression
(further details can be found in the supplementary
material, Ref.~\cite{JeAdDe2017suppl}), including retardation,
for the electrodynamic interaction between two atoms
$A$ and $B$ in arbitrary states,
\begin{multline}
\label{DeltaE}
\Delta E = 
\frac\ii\hbar \int_{0}^\infty \frac{\dd \omega}{2 \pi} \,
\omega^4 \, D_{ij}(\omega, \vec R) \, D_{k\ell}(\omega, \vec R) \, \\
\times\left[\alpha_{A,ik}(\omega) \, 
\alpha_{B,j\ell}(\omega)\pm
\alpha_{\underline{A}B,ik}(\omega) \, 
\alpha_{A\underline{B},j\ell}(\omega)\right] \,,
\end{multline}
where the last term describes the mixing
and is present only for identical atoms.
The photon propagator (in the temporal gauge) 
and the tensor polarizabilities are given by
\allowdisplaybreaks
\begin{subequations}
\begin{align}
D_{ij}(\omega, \vec R) =& \frac{\hbar}{4\pi\epsilon_0 \, c^2}
\left[ \alpha_{ij} + \beta_{ij} \, f(\omega, R) \right]
\frac{\ee^{\ii \frac{\sqrt{\omega^2 + \ii \, \epsilon}}{c}R}}{R} ,
\label{eq:DAlphaBeta}
\\[0.0133ex]
\alpha_{A,ij}(\omega)
=& \; \sum_{v}
\left(
\frac{ \left< \psi_A \left| d_{Ai} \right| v_A \right> \,
\left< v_A \left| d_{Aj} \right| \psi_A \right> }%
{E_{v,A} - \hbar \omega -\ii \epsilon} 
\right.
\nonumber\\[0.0133ex]
& \; \left. +
\frac{ \left< \psi_A \left| d_{Ai} \right| v_A \right> \,
\left< v_A \left| d_{Aj} \right| \psi_A \right> }%
{E_{v,A} + \hbar \omega -\ii \epsilon}
\right) \,,
\label{eq:Pol}\\[0.0133ex]
\alpha_{\underline{A}B,ij}(\omega)
=& \; \sum_{v}
\left(
\frac{ \left< \psi_A \left| d_{Ai} \right| v_A \right> \,
\left< v_B \left| d_{Bj} \right| \psi_B \right> }%
{E_{v,A} - \hbar \omega -\ii \epsilon} 
\right.
\nonumber\\[0.0133ex]
& \; \left. +
\frac{ \left< \psi_A \left| d_{Ai} \right| v_A \right> \,
\left< v_B \left| d_{Bj} \right| \psi_B \right> }%
{E_{v,A} + \hbar \omega -\ii \epsilon}
\right) \,.\label{eq:MixedPol}
\end{align}
\end{subequations}
Here, $f(\omega, R) = \frac{\ii c}{|\omega| R} - \frac{c^2}{\omega^2R^2}$,
and the tensor structures are 
$\alpha_{ij} = \delta_{ij} - \frac{R_i \, R_j}{R^2}$
and $\beta_{ij} = \delta_{ij} - 3 \frac{R_i \, R_j}{R^2}$.
The speed of light is $c$, and $\epsilon_0$ is the vacuum permittivity.
The (excited) state of atom $A$ is $\left|\psi_A\right>$,
and $\vec{d}_A$ is the electric dipole operator for the same atom.
We also write $E_{v,A}\equiv E_v-E_A$ and $E_{v,B}\equiv E_v-E_B$.
As usual, the dipole polarizability is given by a sum
over all virtual states of atom $A$ which are accessible
from $\left|\psi_A\right>$ through an electric dipole transition.
The tensor polarizability 
$\alpha_{A\underline{B},ij}(\omega)$ is obtained from
$\alpha_{\underline{A}B,ij}(\omega)$ by a replacement 
$E_{v,A} \to E_{v,B}$ in the propagator denominators.
For excited reference states,
it is crucial that the polarizabilities~\eqref{eq:Pol}
and~\eqref{eq:MixedPol} have the poles placed according to the 
Feynman prescription; this follows from the time-ordered 
dipole operators which naturally occur 
in time-ordered products of the interaction
Hamiltonian in the $S$ matrix.

If atom $A$ is in an excited state and $B$ in the ground state, 
then the interaction energy $\Delta E = \calQ + \cal W$
[see~\eqref{DeltaE}] can be split 
into a Wick-rotated term ($\omega \to \ii \omega$)
\begin{align}
\label{WICK}
\calW =& \; -\frac{1}{\hbar}\int_{0}^\infty \frac{\dd \omega}{2 \pi} \,
\omega^4 \, D_{ij}(\ii \omega, \vec R) \,
D_{k\ell}(\ii \omega, \vec R) \,
\\[0.0133ex]
& \; \times
[\alpha_{A,ik}(\ii \omega) \, \alpha_{B,j\ell}(\ii \omega) 
\pm \alpha_{A\underline{B},ik}(\ii \omega) \, 
\alpha_{\underline{A}B,j\ell}(\ii \omega) ] \,,
\nonumber
\end{align}
and a pole term from the residues at $\omega = -E_{m,A}/\hbar + \ii \epsilon$,
\begin{subequations}
\begin{widetext}
\label{PresALPHA}
\begin{align}
\label{PresALPHA1}
\calQ =& \; \sum_{E_{m,A} < 0}
\frac{\left< \psi_A \left| d_{Ai} \right| m_A \right> }{(4 \pi \epsilon_0)^2\,R^6} 
\left[ \left< m_A \left| d_{Ak} \right| \psi_A \right> 
\alpha_{B,j\ell}\left(\frac{E_{m,A}}{\hbar}\right) \pm
\left< m_A \left| d_{Bk} \right| \psi_B \right> 
\alpha_{A\underline{B},j\ell}\left( \frac{E_{m,A}}{\hbar} \right)
\right] f_{ijk\ell}(r_{m,A}) \,,
\\[0.1133ex]
\label{PresALPHA2}
f_{ijk\ell}(r) =& \;
-\exp(-2 \ii r) \, \left[ 
\beta_{ij} \, \beta_{k\ell} \, \left( 1
+ 2 \ii \, r \right)
- (2 \alpha_{ij} \, \beta_{k\ell} + \beta_{ij} \, \beta_{k \ell}) r^2
- 2 \ii \alpha_{ij} \, \beta_{k\ell} \, r^3
+ \alpha_{ij} \, \alpha_{k\ell} \, r^4 
\right] \,,
\\[0.1133ex]
\label{PresALPHA3}
{\rm Re} f_{ijk\ell}(r) =& \;
-\cos\left(2 r \right)
\left[  \beta_{ij} \, \beta_{k\ell}
-\left( 2 \alpha_{ij} \, \beta_{k\ell}
+ \beta_{ij} \, \beta_{k\ell} \right) \, r^2
+ \alpha_{ij} \, \alpha_{k\ell} \, r^4
\right]
- 2 r \, \sin\left( r \right)
\left[ \beta_{ij} \, \beta_{k\ell} -
\alpha_{ij} \, \beta_{k\ell} \, r^2 
\right] \,,
\\[0.1133ex]
\label{PresALPHA4}
{\rm Im} f_{ijk\ell}(r) =& \;
-\tfrac12 \, 
\left\{
-2 \sin\left(2 r \right)
\left[  \beta_{ij} \, \beta_{k\ell}
-\left( 2 \alpha_{ij} \, \beta_{k\ell}
+ \beta_{ij} \, \beta_{k\ell} \right) \, r^2
+ \alpha_{ij} \, \alpha_{k\ell} \, r^4
\right]
+ 4 r \, \cos\left( r \right)
\left[ \beta_{ij} \, \beta_{k\ell} -
\alpha_{ij} \, \beta_{k\ell} \, r^2
\right] 
\right\} 
\,,
\end{align}
\end{widetext}
\end{subequations}
where
\begin{equation}
r_{m,A} = \frac{E_{m,A} \, R}{\hbar c} \,,
\qquad
E_{m,A}\equiv E_{m_A}-E_A \,.
\end{equation}
Here, the sum is taken over all states $m$ that are accessible from
$\left|\psi_A\right>$ by a dipole transition \emph{and} of lower energy than
$\left|\psi_A\right>$. 
For a general atom, the generalization is 
trivial: one simply sums the dipole operators of atom 
$A$ over the electrons.

The pole term induces both a real, oscillatory,
distance-dependent energy shift as well as a correction
to the width of the excited state,
\begin{equation}
\calQ = \calP - \frac{\ii}{2} \Gamma \,,
\end{equation}
where $\Gamma$ is obtained from Eq.~\eqref{PresALPHA1}
by substituting for $f_{ijk\ell}(r)$ the 
expression in curly brackets in Eq.~\eqref{PresALPHA4}.

\begin{figure} [t]
\begin{center}
\begin{minipage}{0.99\linewidth}
\begin{center}
\includegraphics[width=0.81\linewidth]{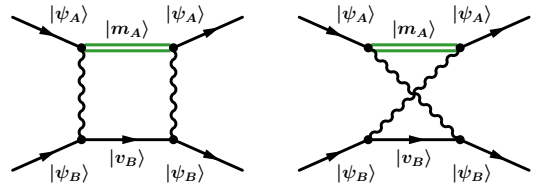}
\caption{\label{fig1} 
Ladder and crossed-ladder Feynman diagrams
for the long-range interaction of atoms. The 
virtual state of atom $A$, labeled $| m_A \rangle$,
is assumed to have a lower energy than the reference state.}
\end{center}
\end{minipage}
\end{center}
\end{figure}

From a QED point of view, the real part of the 
energy shift corresponding to the
pole term is due to a very peculiar process, namely,
resonant virtual emission into vacuum modes
whose angular frequency matches the resonance 
condition $\omega = -E_{m,A} = | E_{m,A} |$.
The resonant emission is accompanied by resonant 
absorption, and therefore leads to a real rather 
than imaginary energy shift. In the ladder and 
crossed-ladder Feynman diagrams (see Fig.~\ref{fig1}),
the virtual electron line of atom $A$, in state $|\psi_A\rangle$,
would turn into a resonant lower-lying virtual state, 
whereas the ground-state atom line is excited into a 
``normal'' energetically higher virtual state $|v_B\rangle$.
The imaginary part of the pole term describes a 
process where the 
virtual photon becomes real and is emitted by the 
atom, in analogy to 
the imaginary part of the self energy~\cite{Be1947,BaSu1978}.
Feynman propagators allow us to reduce the calculation
to only two Feynman diagrams, which capture all possible
time orderings (in contrast to time-ordered perturbation theory).

In Ref.~\cite{DoGuLa2015},
a situation of two non-identical atoms
is considered, with mutually close
resonance energies $\hbar\omega_A$ and $\hbar\omega_B$.
Setting $E_{m,A} = -\hbar\omega_A $ and $E_{q,B} = \hbar\omega_B$,
the authors of Ref.~\cite{DoGuLa2015} assume that
$\omega_A \approx \omega_B$,
and define $\Delta_{AB} = \hbar\omega_A - \hbar\omega_B$
with $| \Delta_{AB} | \ll \hbar\omega_A, \hbar\omega_B$.
Furthermore, they restrict the sum over virtual
states in Eq.~\eqref{PresALPHA} to the resonant state only,
and they only keep the term $1/(E_{m,A} + E_{q,B})$ in
the sum over virtual states, in the polarizability 
$\alpha_B(E_{m,A}/\hbar)$ [see
Eq.~\eqref{PresALPHA}]. Under their assumptions
[see Eq.~(4) of Ref.~\cite{DoGuLa2015}],
$\left| 1/(E_{m,A} + E_{q,B}) \right| =
\left| -1/\Delta_{AB} \right| \gg
\left| 1/(E_{q,B} - E_{m,A}) \right| \approx 1/(2\hbar\omega_B)$.

Under the restriction to the 
resonant virtual states, 
the direct term in Eq.~\eqref{PresALPHA} [proportional to 
$\alpha_{B,j\ell}\left(E_{m,A}/\hbar\right)$]
matches that reported in Ref.~\cite{DoGuLa2015}
if we average the latter over the interaction time
$T > 2 R/c$.  Our result adds the 
contribution from nonresonant virtual 
states, which allow us to match the 
result with the close-range (\vdw{}) limit, 
as well as the mixing term [proportional to 
$\alpha_{A\underline{B},j\ell}\left( E_{m,A}/\hbar \right)$]
and the imaginary part of the energy shift (width term).
For the mixing term to be nonzero, we need
the orbital angular momentum quantum numbers to fulfill
the relation $\ell_A=\ell_B$ or $\left|\ell_A-\ell_B\right|=2$, by virtue of the
usual selection rules of atomic physics.
Furthermore, we find that the full consideration of the 
Wick-rotated term and the pole term is crucial for 
obtaining numerically correct results for the 
interaction energies, for surprisingly large interatomic 
distances.

{\em Numerical Calculations.---}In the following, 
we aim to apply the developed formalism
to $nD$--$1S$ atomic hydrogen systems. 
The interaction energy depends both on the 
spin orientation of the electron (total angular 
momentum $J$) as well as its projection $\mu$
onto the quantization axis~\cite{JeAdDe2017suppl}.
One may eliminate this dependence
by evaluating the average over the fine-structure 
resolved states. 

{\em Short Range.---}For interatomic separations in the
range $a_0\ll R\ll a_0/\alpha$ (where $a_0$ is the Bohr radius),
the interaction energy \eqref{DeltaE} is well approximated as
$\Delta E \approx \Delta E_{\rm vdW}$ where
\begin{align}
\label{VDWres} 
\Delta E_{\rm vdW} = & \; -
\frac{1}{(4 \pi \epsilon_0)^2}
\frac{\beta_{ij} \, \beta_{k\ell}}{R^6} \,
\sum_{v} \sum_{q}
\frac{1}{E_{v,A} + E_{q,B}} 
\nonumber\\[0.0133ex]
&\; \times \left[ \left< \psi_A \left| d_{Ai} \right| v_A \right>
\left< v_A \left| d_{Ak} \right| \psi_A  \right>\right.
\nonumber\\[0.0133ex]
&\hspace{50pt}\times \left< \psi_B \left| d_{Bj} \right| q_B \right>
\left< q_B \left| d_{B\ell} \right| \psi_B \right>
\nonumber\\[0.0133ex]
&\pm\left< \psi_A \left| d_{Ai} \right| v_A \right>
\left< v_B \left| d_{Bk} \right| \psi_B  \right>
\nonumber\\[0.0133ex]
&\left.\hspace{50pt}\times \left< \psi_B \left| d_{Bj} \right| q_B \right>
\left< q_A \left| d_{A\ell} \right| \psi_A \right>\right] \,,
\nonumber\\[0.0133ex]
= & \; -\frac{1}{R^6}\left(D_6\left(A;B\right)\pm
M_6\left(A;B\right)\right) \,.
\end{align}
Here, $D_6$ is the direct, and $M_6$ is the mixing \vdw{}
coefficient~\cite{Ch1972,DeYo1973,AdEtAl2017vdWi,JeEtAl2017vdWii,%
PoTh1995,SaBuWeDu2006}. For energetically lower states
in atom $A$ (with $E_{v,A} = E_{m,A} < 0$), the representation~\eqref{VDWres}
is obtained by carefully considering the contributions from the
Wick-rotated term $\calW$ and the pole term $\calP$.

For the fine-structure average of the 
direct term $D_6$, we have
\begin{equation} 
\langle D_6(nD;1S) \rangle =
\langle D_6^{\left(P\right)}(nD;1S) \rangle +
\langle D_6^{\left(F\right)}(nD;1S) \rangle \,,
\end{equation}
where the virtual-$P$-state contribution 
$\langle D_6^{\left(P\right)}(nD;1S) \rangle$
and the virtual-$F$-state contribution 
$\langle D_6^{\left(F\right)}(nD;1S) \rangle$
are given in Table~\ref{table1}.
Numerically, we find that the mixing term $M_6$ is smaller
than the direct term $D_6$, by at least four orders of
magnitude, for all fine-structure resolved $nD$ states,
for all distance ranges investigated in this Letter.
This trend follows the pattern observed
for the \vdw{} coefficients (Table~\ref{table1})
and is in contrast to the $2S$--$1S$ system,
where both terms are of comparable magnitude~\cite{Ch1972,AdEtAl2017vdWi}.

\begin{table} [t]
\caption{\label{table1}
\Vdw{} $D_6$ (direct) coefficients, for $nD$--$1S$
interactions, averaged over the total angular
momenta and magnetic projections of the excited $D$ state.
The coefficients are given in units of $E_h\,a_0^6$.}
\begin{center}
\begin{tabular}{l@{\hspace{3mm}}r@{\hspace{3mm}}r@{\hspace{3mm}}r}
\hline
\hline
\stdrule
Coefficient & Virtual $P$  & Virtual $F$ & Total \\
\hline
\stdrule
$\langle D_6(8D; 1S)  \rangle$ &  17459.439 &  26156.866  & 43616.296 \\
\stdrule
$\langle D_6(10D; 1S) \rangle$ &  43476.563 &  65182.580  & 108659.144 \\
\stdrule
$\langle D_6(12D; 1S) \rangle$ &  91115.328 & 136640.733  & 227756.061 \\
\hline
\hline
\end{tabular}
\end{center}
\end{table}

\begin{figure} [t]
\begin{center}
\begin{minipage}{0.99\linewidth}
\begin{center}
\includegraphics[width=0.81\linewidth]{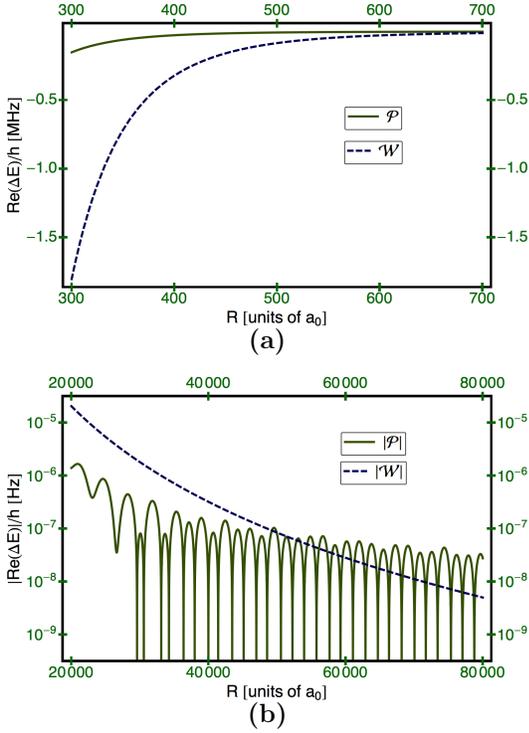}
\caption{\label{fig2} 
The upper panel (a) compares the Wick-rotated term $\calW$
(dashed line) to the real part $\calP = {\rm Re} \, \calQ$ of the pole term 
(solid line) for the 
$12D$--$1S$ interaction (fine-structure average) in the 
intermediate range $a_0/\alpha \lesssim R \ll \hbar c/\calL$.
The interaction energy is $\Delta E = \calQ + \calW$.
The Wick-rotated term dominates despite parametric
suppression [see Eq.~\eqref{param}].
The lower panel~(b) displays the logarithm of the 
modulus of the interaction energy (Wick-rotated term versus 
pole term) for very large interatomic distance;
the pole term finally dominates.
The oscillations of the pole term are dominated by virtual $2P$ 
state contribution; when the interaction energy 
changes sign, the logarithm diverges to $-\infty$.}
\end{center}
\end{minipage}
\end{center}
\end{figure}

{\em Long Range.---}One might think that the 
$1/R^2$ pole term from Eq.~\eqref{PresALPHA}
should easily dominate the interaction energy 
in the range $R \gg a_0/\alpha$. Indeed, power counting 
in the fine-structure constant $\alpha$,
according to Ref.~\cite{BeSa1957}, shows that 
the cosine and sine terms in $\calP$ 
are asymptotically given by
\begin{subequations}
\label{param}
\begin{align}
\label{param1}
\frac{\calP_{\rm cos}}{E_h} \sim & \;
\left\{ 
\frac{\alpha^4}{\rho^2} \cos(\alpha \, \rho) \,,
\frac{\alpha^2}{\rho^4} \cos(\alpha \, \rho) \,,
\frac{1}{\rho^6} \cos(\alpha \, \rho) 
\right\} \,,
\\[0.1133ex]
\label{param2}
\frac{\calP_{\rm sin}}{E_h} \sim & \;
\left\{
\frac{\alpha^3}{\rho^3} \sin(\alpha \, \rho) \,,
\frac{\alpha}{\rho^5} \sin(\alpha \, \rho) 
\right\} \,,
\end{align}
where $\rho = R/a_0$ and $E_h$ is the 
Hartree energy. A comparison to the \vdw{} term,
given in Eq.~\eqref{VDWres},
and the Wick-rotated term~\eqref{WICK},
\begin{equation}
\label{param3}
\Delta E_{\rm vdW} \sim \frac{E_h}{\rho^6} \,,
\qquad
\calW \; \mathop{\sim}^{\rho \gg 1/\alpha} \; \frac{E_h}{\alpha \, \rho^7} \,,
\end{equation}
\end{subequations}
shows that all terms (pole term, Wick-rotated, and 
\vdw{}) are of the same order-of-magnitude for 
$\rho \sim 1/\alpha$, while
the pole term should parametrically dominate 
for $\rho > 1/\alpha$.
However, this consideration does not 
take into account the scaling of the 
terms with the principal quantum number $n$.
While we find that the $D_6$ coefficients 
typically grow as $n^4$ for a given manifold of 
states (see also~Ref.~\cite{AdEtAl2017vdWiii}),
the energy differences $E_{m,A}$ for 
adjacent lower-lying states 
are proportional to $1/(n-1)^2 - 1/n^2 \sim 1/n^3$ 
for large $n$, and the fourth power of the
energy difference $E_{m,A}$ enters the prefactor 
of the $1/R^2$ pole term. Hence, 
it is interesting to compare the parametric
estimates to a concrete calculation 
for excited $nD$ states; this is also important 
in order to gauge the importance of the nonresonant 
contributions to the interaction energy
which were left out in Ref.~\cite{DoGuLa2015}.

But let us first write down the leading asymptotic terms,
for all long-range contributions of interest.
For $R \gg \hbar c/\calL$, 
where $\calL$ is the Lamb shift energy
of about $1 \, {\rm GHz}$ (see Ref.~\cite{AdEtAl2017vdWi}),
the Wick-rotated contribution attains the 
familiar $1/R^7$ asymptotics from the 
Casimir--Polder formalism~\cite{CaPo1948},
\begin{equation}
\label{LONGWICK_GdXc}
\calW^{\left(\mathrm{dir}\right)} 
\mathop{=}^{R \to \infty} 
-\frac{\hbar c}{8\pi}
\frac{\alpha_{nD,ij}(0) \alpha_{1S}(0)}%
{ (4 \pi \epsilon_0)^2 \, R^7}
\left(13\,\delta_{ij}+15\frac{R_iR_j}{R^2}\right) \,.
\end{equation}
This tail is parametrically 
suppressed in comparison to 
the leading $1/R^2$ pole contribution,
\begin{multline}
\label{LONGPOLE_GdXc}
\calP^{\left(\mathrm{dir}\right)}  \mathop{=}^{R \to \infty}
-\sum_{E_m<E_{nD}}
\left(\frac{E_{m,A}}{\hbar c} \right)^4 \,
\frac{\cos\left(2\frac{E_{m,A}}{\hbar c}R\right)}{(4 \pi \epsilon_0)^2 \,R^2} \,
\\[0.0133ex]
\times \alpha_{ij} \left< nD \left| d_{Ai} \right| m_A \right>
\left< m_A \left| d_{Aj} \right| nD \right> \,
\alpha_{1S}\left(\frac{E_{m,A}}{\hbar}\right)\,.
\end{multline}
In the intermediate range $a_0 \ll R \ll \hbar c/\calL$,
the Wick-rotated contribution has a nonretarded 
$1/R^6$ tail, due to a nonretarded contribution from
virtual $nP$ and $nF$ states
which are displaced from the $nD$ state only 
by the Lamb shift [see Eqs.~(23) and~(24) of 
Ref.~\cite{AdEtAl2017vdWi}],
\begin{equation}
\calW^{\left(\mathrm{dir}\right)} \sim
- \frac{\overline D_6(nD;1S)}{R^6} \,,
\qquad
\frac{a_0}{\alpha} \ll R \ll \frac{\hbar c}{\calL} \,.
\end{equation}
The fine-structure average
of $\overline D_6(nD;1S)$ is given by~\cite{JeAdDe2017suppl}
\begin{equation}
\label{D6bar}
\langle \overline D_6(nD;1S) 
\rangle = \frac{81}{8} n^2 \, (n^2 - 7) \,.
\end{equation}
For the mixing term, simplifications are scarce; the 
leading long-range asymptotics of the 
Wick-rotated term read
\begin{multline}
\label{LONGMWICK_GdXc}
\calW^{\left(\mathrm{mix}\right)}  
\mathop{=}^{R \to \infty} \;
-\frac{\hbar c}{8\pi}
\frac{ \alpha_{\underline{nD}\,1S,ik}(0) \,
\alpha_{nD\,\underline{1S},j\ell}(0) }{ (4 \pi \epsilon_0)^2 \, R^7}
\\
\times \left( 3 \alpha_{ij}\alpha_{kl} 
+ 5 \alpha_{ij}\beta_{kl} 
+ 5 \beta_{ij}\beta_{kl} \right) \,.
\end{multline}
The leading pole contribution (in the long range) is
given by a sum over virtual $P$ states
which enter the mixed polarizability
$\alpha_{nD\,\underline{1S},j\ell}$,
\begin{multline}
\label{LONGMPOLE_GdXct}
\calP^{\left(\mathrm{mix}\right)}
\mathop{=}^{R \to \infty}
-\frac{\sum_{E_m<E_{nD}}}{(4 \pi \epsilon_0)^2}
\left(\frac{E_{m,A}}{\hbar c} \right)^4
\frac{\cos\left(2\frac{E_{m,A}}{\hbar c}R\right)}{R^2}
\\[0.0133ex]
\times\alpha_{ij}\,\alpha_{kl}\,
\alpha_{nD\,\underline{1S},j\ell}\left(\frac{E_{m,A}}{\hbar}\right)
\\[0.1133ex]
\times\,\left< nD \left| d_{Ai} \right| m_A \right>
\left< m_A \left| d_{Ak} \right| 1S \right> \,.
\end{multline}
In the intermediate range, one has
\begin{equation}
\calW^{\left(\mathrm{mix}\right)} \sim
- \frac{\overline M_6(nD;1S)}{R^6} \,,
\qquad
\frac{a_0}{\alpha} \ll R \ll \frac{\hbar c}{\calL} \,,
\end{equation}
where $\overline M_6(nD;1S)$ is the generalization 
of $\overline D_6$ to the mixing term
[see Ref.~\cite{JeAdDe2017suppl} and
Eq.~(67) of Ref.~\cite{AdEtAl2017vdWi}].

In Fig.~\ref{fig2}, we compare the magnitude of the 
Wick-rotated term and the pole term in the intermediate 
range $a_0/\alpha \ll R \ll \hbar c/\calL$, and for 
very large separations $R \sim \hbar c/\calL$.
While a parametric analysis [Eq.~\eqref{param}]
would suggest dominance of the pole term in the intermediate 
range, a numerical calculation reveals a different behavior, 
with the Wick-rotated term dominating the interaction,
due to the variability of the numerical coefficients
multiplying the parametric estimates given in Eq.~\eqref{param}.

%
%

{\em Conclusions.---}We have shown that 
the consistent use of Feynman propagators 
and the concomitant virtual photon integration contours lead
to the prediction of long-range tails for excited-state \vdw{} 
interactions. Pole
terms are picked up for virtual states
$\left|v_A\right> = \left|m_A\right>$ of lower
energy than the reference state of the excited atom $A$. 
The pole contribution $\calQ$ to the energy 
shift is complex rather than real (includes a width term
$\Gamma = -2 \, {\rm Im} \calQ$),
is spacewise-oscillating and in the long-range, behaves as
$\cos[2 (E_m-E_A) \, R/(\hbar c)]/R^2$, where $E_A$ is the reference state
energy and $E_m<E_A$ that of the low-energy virtual state.
For excited states, both the direct as well as the exchange
(gerade-ungerade mixing) term
can be expressed as a sum of a Wick-rotated 
contribution [Eq.~\eqref{WICK}],
and a pole term [Eq.~\eqref{PresALPHA}].
Our inclusion of the nonresonant terms in the interaction energy
enables us to match the very-long-range, oscillatory result
against the well-known close-range, nonretarded van der Waals limit, 
and to carry out numerical calculations in the intermediate region.
We also include the width term, and the gerade-ungerade mixing term which pertains to 
excited-state interactions of identical atoms.

For $nD$--$1S$ interactions,
we have shown that despite parametric suppression,
the Wick-rotated term, which is non-oscillatory
and contains the non-resonant states,
still dominates in the intermediate distance 
range $a_0/\alpha \lesssim R \ll \hbar c/\calL$
(see Fig.~\ref{fig2}).
The very-long-range, oscillatory tail of the \vdw{} interaction 
is relevant only for very large interatomic distances.
This conclusion holds for $nD$--$1S$ interactions
as well as $nS$--$1S$ systems~\cite{JeAdDe2017suppl,AdEtAl2017vdWiii}. 
The reason for the suppression is that the 
numerical coefficients which multiply the 
parametric estimates given in 
Eq.~\eqref{param} drastically depend on the 
particular term in the \vdw{} energy.
This is in part due to the scaling of the
coefficients with the principal quantum number.
E.g., for $nD$--$1S$ interactions, the $1/R^2$ leading 
oscillatory tail from Eq.~\eqref{PresALPHA} 
is of order $E_h \, \alpha^4 \, \cos(\alpha \, \rho) / \rho^2$,
yet multiplied by numerical coefficients of order $10^{-6}$
[in addition to the factor $\alpha^4$; see
the supplementary material~\cite{JeAdDe2017suppl},
Eq.~\eqref{D6bar}, Table~\ref{table1} and Fig.~\ref{fig2}].
By contrast, the non-oscillatory terms of 
order $E_h/\rho^6$ are multiplied by 
coefficients of order $10^4 \ldots 10^6$.
This behavior of the coefficients
changes any predictions based on 
the parametric estimates given in Eq.~\eqref{param} 
by ten orders of magnitude as compared to 
a situation with coefficients of order unity.

Our results are important for an improved 
analysis of pressure shifts, and 
fluctuating-dipole-induced energy shift, 
for atomic beam spectroscopy with Rydberg states,
where these effects have been identified as notoriously 
problematic in recent years
(see pp.~134 and 151 of Ref.~\cite{dV2002}).
An improved determination of the Rydberg constant
based on Rydberg-state spectroscopy could 
resolve the muonic hydrogen proton radius puzzle,
because the smaller proton radius measured 
in Ref.~\cite{PoEtAl2010CREMA,PoEtAl2016CREMA} leads to 
a Rydberg constant which is discrepant with regard
to the current CODATA value~\cite{PoEtAl2010CREMA,MoNeTa2016}.

{\em Acknowledgments.---}This research has been supported 
by the NSF (grant PHY--1403937).


\begin{thebibliography}{26}%
\makeatletter
\providecommand \@ifxundefined [1]{%
 \@ifx{#1\undefined}
}%
\providecommand \@ifnum [1]{%
 \ifnum #1\expandafter \@firstoftwo
 \else \expandafter \@secondoftwo
 \fi
}%
\providecommand \@ifx [1]{%
 \ifx #1\expandafter \@firstoftwo
 \else \expandafter \@secondoftwo
 \fi
}%
\providecommand \natexlab [1]{#1}%
\providecommand \enquote  [1]{``#1''}%
\providecommand \bibnamefont  [1]{#1}%
\providecommand \bibfnamefont [1]{#1}%
\providecommand \citenamefont [1]{#1}%
\providecommand \href@noop [0]{\@secondoftwo}%
\providecommand \href [0]{\begingroup \@sanitize@url \@href}%
\providecommand \@href[1]{\@@startlink{#1}\@@href}%
\providecommand \@@href[1]{\endgroup#1\@@endlink}%
\providecommand \@sanitize@url [0]{\catcode `\\12\catcode `\$12\catcode
  `\&12\catcode `\#12\catcode `\^12\catcode `\_12\catcode `\%12\relax}%
\providecommand \@@startlink[1]{}%
\providecommand \@@endlink[0]{}%
\providecommand \url  [0]{\begingroup\@sanitize@url \@url }%
\providecommand \@url [1]{\endgroup\@href {#1}{\urlprefix }}%
\providecommand \urlprefix  [0]{URL }%
\providecommand \Eprint [0]{\href }%
\providecommand \doibase [0]{http://dx.doi.org/}%
\providecommand \selectlanguage [0]{\@gobble}%
\providecommand \bibinfo  [0]{\@secondoftwo}%
\providecommand \bibfield  [0]{\@secondoftwo}%
\providecommand \translation [1]{[#1]}%
\providecommand \BibitemOpen [0]{}%
\providecommand \bibitemStop [0]{}%
\providecommand \bibitemNoStop [0]{.\EOS\space}%
\providecommand \EOS [0]{\spacefactor3000\relax}%
\providecommand \BibitemShut  [1]{\csname bibitem#1\endcsname}%
\let\auto@bib@innerbib\@empty
\bibitem [{\citenamefont {Casimir}\ and\ \citenamefont
  {Polder}(1948)}]{CaPo1948}%
  \BibitemOpen
  \bibfield  {author} {\bibinfo {author} {\bibfnamefont {H.~B.~G.}\
  \bibnamefont {Casimir}}\ and\ \bibinfo {author} {\bibfnamefont
  {D.}~\bibnamefont {Polder}},\ }\bibfield  {title} {\enquote {\bibinfo {title}
  {\relax{The Influence of Radiation on the London-van-der-Waals Forces}},}\
  }\href@noop {} {\bibfield  {journal} {\bibinfo  {journal} {Phys. Rev.}\
  }\textbf {\bibinfo {volume} {73}},\ \bibinfo {pages} {360--372} (\bibinfo
  {year} {1948})}\BibitemShut {NoStop}%
\bibitem [{\citenamefont {Chibisov}(1972)}]{Ch1972}%
  \BibitemOpen
  \bibfield  {author} {\bibinfo {author} {\bibfnamefont {M.~I.}\ \bibnamefont
  {Chibisov}},\ }\bibfield  {title} {\enquote {\bibinfo {title}
  {\relax{Dispersion Interaction of Neutral Atoms}},}\ }\href@noop {}
  {\bibfield  {journal} {\bibinfo  {journal} {Opt. Spectrosc.}\ }\textbf
  {\bibinfo {volume} {32}},\ \bibinfo {pages} {1--3} (\bibinfo {year}
  {1972})}\BibitemShut {NoStop}%
\bibitem [{\citenamefont {Deal}\ and\ \citenamefont {Young}(1973)}]{DeYo1973}%
  \BibitemOpen
  \bibfield  {author} {\bibinfo {author} {\bibfnamefont {W.~J.}\ \bibnamefont
  {Deal}}\ and\ \bibinfo {author} {\bibfnamefont {R.~H.}\ \bibnamefont
  {Young}},\ }\bibfield  {title} {\enquote {\bibinfo {title}
  {\relax{Long--Range Dispersion Interactions Involving Excited Atoms; the
  H(1s)---H(2s) Interaction}},}\ }\href@noop {} {\bibfield  {journal} {\bibinfo
   {journal} {Int. J. Quantum Chem.}\ }\textbf {\bibinfo {volume} {7}},\
  \bibinfo {pages} {877--892} (\bibinfo {year} {1973})}\BibitemShut {NoStop}%
\bibitem [{\citenamefont {Adhikari}\ \emph {et~al.}(2017)\citenamefont
  {Adhikari}, \citenamefont {Debierre}, \citenamefont {Matveev}, \citenamefont
  {Kolachevsky},\ and\ \citenamefont {Jentschura}}]{AdEtAl2017vdWi}%
  \BibitemOpen
  \bibfield  {author} {\bibinfo {author} {\bibfnamefont {C.~M.}\ \bibnamefont
  {Adhikari}}, \bibinfo {author} {\bibfnamefont {V.}~\bibnamefont {Debierre}},
  \bibinfo {author} {\bibfnamefont {A.}~\bibnamefont {Matveev}}, \bibinfo
  {author} {\bibfnamefont {N.}~\bibnamefont {Kolachevsky}}, \ and\ \bibinfo
  {author} {\bibfnamefont {U.~D.}\ \bibnamefont {Jentschura}},\ }\bibfield
  {title} {\enquote {\bibinfo {title} {\relax{Long-range interactions of
  hydrogen atoms in excited states.~I. $2S$--$1S$ interactions and
  Dirac--$\delta$ perturbations}},}\ }\href@noop {} {\bibfield  {journal}
  {\bibinfo  {journal} {Phys. Rev. A}\ }\textbf {\bibinfo {volume} {95}},\
  \bibinfo {pages} {022703} (\bibinfo {year} {2017})}\BibitemShut {NoStop}%
\bibitem [{\citenamefont {Jentschura}\ \emph
  {et~al.}(2017{\natexlab{a}})\citenamefont {Jentschura}, \citenamefont
  {Debierre}, \citenamefont {Adhikari}, \citenamefont {Matveev},\ and\
  \citenamefont {Kolachevsky}}]{JeEtAl2017vdWii}%
  \BibitemOpen
  \bibfield  {author} {\bibinfo {author} {\bibfnamefont {U.~D.}\ \bibnamefont
  {Jentschura}}, \bibinfo {author} {\bibfnamefont {V.}~\bibnamefont
  {Debierre}}, \bibinfo {author} {\bibfnamefont {C.~M.}\ \bibnamefont
  {Adhikari}}, \bibinfo {author} {\bibfnamefont {A.}~\bibnamefont {Matveev}}, \
  and\ \bibinfo {author} {\bibfnamefont {N.}~\bibnamefont {Kolachevsky}},\
  }\bibfield  {title} {\enquote {\bibinfo {title} {\relax{Long-range
  interactions of excited hydrogen atoms. II. Hyperfine-resolved $2S$--$2S$
  system}},}\ }\href@noop {} {\bibfield  {journal} {\bibinfo  {journal} {Phys.
  Rev. A}\ }\textbf {\bibinfo {volume} {95}},\ \bibinfo {pages} {022704}
  (\bibinfo {year} {2017}{\natexlab{a}})}\BibitemShut {NoStop}%
\bibitem [{\citenamefont {Power}\ and\ \citenamefont
  {Thirunamachandran}(1995)}]{PoTh1995}%
  \BibitemOpen
  \bibfield  {author} {\bibinfo {author} {\bibfnamefont {E.~A.}\ \bibnamefont
  {Power}}\ and\ \bibinfo {author} {\bibfnamefont {T.}~\bibnamefont
  {Thirunamachandran}},\ }\bibfield  {title} {\enquote {\bibinfo {title}
  {\relax{Dispersion forces between molecules with one or both molecules
  excited}},}\ }\href@noop {} {\bibfield  {journal} {\bibinfo  {journal} {Phys.
  Rev. A}\ }\textbf {\bibinfo {volume} {51}},\ \bibinfo {pages} {3660--3666}
  (\bibinfo {year} {1995})}\BibitemShut {NoStop}%
\bibitem [{\citenamefont {Safari}\ \emph {et~al.}(2006)\citenamefont {Safari},
  \citenamefont {Buhmann}, \citenamefont {Welsch},\ and\ \citenamefont
  {Dung}}]{SaBuWeDu2006}%
  \BibitemOpen
  \bibfield  {author} {\bibinfo {author} {\bibfnamefont {H.}~\bibnamefont
  {Safari}}, \bibinfo {author} {\bibfnamefont {S.~Y.}\ \bibnamefont {Buhmann}},
  \bibinfo {author} {\bibfnamefont {D.-G.}\ \bibnamefont {Welsch}}, \ and\
  \bibinfo {author} {\bibfnamefont {H.~T.}\ \bibnamefont {Dung}},\ }\bibfield
  {title} {\enquote {\bibinfo {title} {\relax{Body-assisted van der Waals
  interaction between two atoms}},}\ }\href@noop {} {\bibfield  {journal}
  {\bibinfo  {journal} {Phys. Rev. A}\ }\textbf {\bibinfo {volume} {74}},\
  \bibinfo {pages} {042101} (\bibinfo {year} {2006})}\BibitemShut {NoStop}%
\bibitem [{\citenamefont {Safari}\ and\ \citenamefont
  {Karimpour}(2015)}]{SaKa2015}%
  \BibitemOpen
  \bibfield  {author} {\bibinfo {author} {\bibfnamefont {H.}~\bibnamefont
  {Safari}}\ and\ \bibinfo {author} {\bibfnamefont {M.~R.}\ \bibnamefont
  {Karimpour}},\ }\bibfield  {title} {\enquote {\bibinfo {title}
  {\relax{Body-Assisted van der Waals Interaction between Excited Atoms}},}\
  }\href@noop {} {\bibfield  {journal} {\bibinfo  {journal} {Phys. Rev. Lett.}\
  }\textbf {\bibinfo {volume} {114}},\ \bibinfo {pages} {013201} (\bibinfo
  {year} {2015})}\BibitemShut {NoStop}%
\bibitem [{\citenamefont {Berman}(2015)}]{Be2015}%
  \BibitemOpen
  \bibfield  {author} {\bibinfo {author} {\bibfnamefont {P.~R.}\ \bibnamefont
  {Berman}},\ }\bibfield  {title} {\enquote {\bibinfo {title}
  {\relax{Interaction energy of nonidentical atoms}},}\ }\href@noop {}
  {\bibfield  {journal} {\bibinfo  {journal} {Phys. Rev. A}\ }\textbf {\bibinfo
  {volume} {91}},\ \bibinfo {pages} {042127} (\bibinfo {year}
  {2015})}\BibitemShut {NoStop}%
\bibitem [{\citenamefont {Milonni}\ and\ \citenamefont
  {Rafsanjani}(2015)}]{MiRa2015}%
  \BibitemOpen
  \bibfield  {author} {\bibinfo {author} {\bibfnamefont {P.~W.}\ \bibnamefont
  {Milonni}}\ and\ \bibinfo {author} {\bibfnamefont {S.~M.~H.}\ \bibnamefont
  {Rafsanjani}},\ }\bibfield  {title} {\enquote {\bibinfo {title}
  {\relax{Distance dependence of two-atom dipole interactions with one atom in
  an excited state}},}\ }\href@noop {} {\bibfield  {journal} {\bibinfo
  {journal} {Phys. Rev. A}\ }\textbf {\bibinfo {volume} {92}},\ \bibinfo
  {pages} {062711} (\bibinfo {year} {2015})}\BibitemShut {NoStop}%
\bibitem [{\citenamefont {Donaire}\ \emph {et~al.}(2015)\citenamefont
  {Donaire}, \citenamefont {Gu\'{e}rout},\ and\ \citenamefont
  {Lambrecht}}]{DoGuLa2015}%
  \BibitemOpen
  \bibfield  {author} {\bibinfo {author} {\bibfnamefont {M.}~\bibnamefont
  {Donaire}}, \bibinfo {author} {\bibfnamefont {R.}~\bibnamefont
  {Gu\'{e}rout}}, \ and\ \bibinfo {author} {\bibfnamefont {A.}~\bibnamefont
  {Lambrecht}},\ }\bibfield  {title} {\enquote {\bibinfo {title}
  {\relax{Quasiresonant van der Waals Interaction between Nonidentical
  Atoms}},}\ }\href@noop {} {\bibfield  {journal} {\bibinfo  {journal} {Phys.
  Rev. Lett.}\ }\textbf {\bibinfo {volume} {115}},\ \bibinfo {pages} {033201}
  (\bibinfo {year} {2015})}\BibitemShut {NoStop}%
\bibitem [{\citenamefont {Donaire}(2016)}]{Do2016}%
  \BibitemOpen
  \bibfield  {author} {\bibinfo {author} {\bibfnamefont {M.}~\bibnamefont
  {Donaire}},\ }\bibfield  {title} {\enquote {\bibinfo {title} {\relax{Two-atom
  interaction energies with one atom in an excited state: van der Waals
  potentials versus level shifts}},}\ }\href@noop {} {\bibfield  {journal}
  {\bibinfo  {journal} {Phys. Rev. A}\ }\textbf {\bibinfo {volume} {93}},\
  \bibinfo {pages} {052706} (\bibinfo {year} {2016})}\BibitemShut {NoStop}%
\bibitem [{\citenamefont {Gomberoff}\ \emph {et~al.}(1966)\citenamefont
  {Gomberoff}, \citenamefont {McLone},\ and\ \citenamefont
  {Power}}]{GoMLPo1966}%
  \BibitemOpen
  \bibfield  {author} {\bibinfo {author} {\bibfnamefont {L.}~\bibnamefont
  {Gomberoff}}, \bibinfo {author} {\bibfnamefont {R.~R.}\ \bibnamefont
  {McLone}}, \ and\ \bibinfo {author} {\bibfnamefont {E.~A.}\ \bibnamefont
  {Power}},\ }\bibfield  {title} {\enquote {\bibinfo {title}
  {\relax{Long--Range Retarded Potentials between Molecules}},}\ }\href@noop {}
  {\bibfield  {journal} {\bibinfo  {journal} {J. Chem. Phys.}\ }\textbf
  {\bibinfo {volume} {44}},\ \bibinfo {pages} {4148--4153} (\bibinfo {year}
  {1966})}\BibitemShut {NoStop}%
\bibitem [{\citenamefont {Biraben}\ \emph {et~al.}(1989)\citenamefont
  {Biraben}, \citenamefont {Garreau}, \citenamefont {Julien},\ and\
  \citenamefont {Allegrini}}]{BiGaJuAl1989}%
  \BibitemOpen
  \bibfield  {author} {\bibinfo {author} {\bibfnamefont {F.}~\bibnamefont
  {Biraben}}, \bibinfo {author} {\bibfnamefont {J.-C.}\ \bibnamefont
  {Garreau}}, \bibinfo {author} {\bibfnamefont {L.}~\bibnamefont {Julien}}, \
  and\ \bibinfo {author} {\bibfnamefont {M.}~\bibnamefont {Allegrini}},\
  }\bibfield  {title} {\enquote {\bibinfo {title} {\relax{New Measurement of
  the Rydberg Constant by Two-Photon Spectroscopy of Hydrogen Rydberg
  States}},}\ }\href@noop {} {\bibfield  {journal} {\bibinfo  {journal} {Phys.
  Rev. Lett.}\ }\textbf {\bibinfo {volume} {62}},\ \bibinfo {pages} {621--621}
  (\bibinfo {year} {1989})}\BibitemShut {NoStop}%
\bibitem [{\citenamefont {de~Beauvoir}\ \emph {et~al.}(1997)\citenamefont
  {de~Beauvoir}, \citenamefont {Nez}, \citenamefont {Julien}, \citenamefont
  {Cagnac}, \citenamefont {Biraben}, \citenamefont {Touahri}, \citenamefont
  {Hilico}, \citenamefont {Acef}, \citenamefont {Clairon},\ and\ \citenamefont
  {Zondy}}]{BeEtAl1997}%
  \BibitemOpen
  \bibfield  {author} {\bibinfo {author} {\bibfnamefont {B.}~\bibnamefont
  {de~Beauvoir}}, \bibinfo {author} {\bibfnamefont {F.}~\bibnamefont {Nez}},
  \bibinfo {author} {\bibfnamefont {L.}~\bibnamefont {Julien}}, \bibinfo
  {author} {\bibfnamefont {B.}~\bibnamefont {Cagnac}}, \bibinfo {author}
  {\bibfnamefont {F.}~\bibnamefont {Biraben}}, \bibinfo {author} {\bibfnamefont
  {D.}~\bibnamefont {Touahri}}, \bibinfo {author} {\bibfnamefont
  {L.}~\bibnamefont {Hilico}}, \bibinfo {author} {\bibfnamefont
  {O.}~\bibnamefont {Acef}}, \bibinfo {author} {\bibfnamefont {A.}~\bibnamefont
  {Clairon}}, \ and\ \bibinfo {author} {\bibfnamefont {J.~J.}\ \bibnamefont
  {Zondy}},\ }\bibfield  {title} {\enquote {\bibinfo {title} {\relax{Absolute
  Frequency Measurement of the $2S$--$8S/D$ Transitions in Hydrogen and
  Deuterium: New Determination of the Rydberg Constant}},}\ }\href@noop {}
  {\bibfield  {journal} {\bibinfo  {journal} {Phys. Rev. Lett.}\ }\textbf
  {\bibinfo {volume} {78}},\ \bibinfo {pages} {440--443} (\bibinfo {year}
  {1997})}\BibitemShut {NoStop}%
\bibitem [{\citenamefont {Schwob}\ \emph {et~al.}(1999)\citenamefont {Schwob},
  \citenamefont {Jozefowski}, \citenamefont {de~Beauvoir}, \citenamefont
  {Hilico}, \citenamefont {Nez}, \citenamefont {Julien}, \citenamefont
  {Biraben}, \citenamefont {Acef}, \citenamefont {Zondy},\ and\ \citenamefont
  {Clairon}}]{ScEtAl1999}%
  \BibitemOpen
  \bibfield  {author} {\bibinfo {author} {\bibfnamefont {C.}~\bibnamefont
  {Schwob}}, \bibinfo {author} {\bibfnamefont {L.}~\bibnamefont {Jozefowski}},
  \bibinfo {author} {\bibfnamefont {B.}~\bibnamefont {de~Beauvoir}}, \bibinfo
  {author} {\bibfnamefont {L.}~\bibnamefont {Hilico}}, \bibinfo {author}
  {\bibfnamefont {F.}~\bibnamefont {Nez}}, \bibinfo {author} {\bibfnamefont
  {L.}~\bibnamefont {Julien}}, \bibinfo {author} {\bibfnamefont
  {F.}~\bibnamefont {Biraben}}, \bibinfo {author} {\bibfnamefont
  {O.}~\bibnamefont {Acef}}, \bibinfo {author} {\bibfnamefont {J.~J.}\
  \bibnamefont {Zondy}}, \ and\ \bibinfo {author} {\bibfnamefont
  {A.}~\bibnamefont {Clairon}},\ }\bibfield  {title} {\enquote {\bibinfo
  {title} {\relax{Optical Frequency Measurement of the $2S$-$12D$ Transitions
  in Hydrogen and Deuterium: Rydberg Constant and Lamb Shift
  Determinations}},}\ }\href@noop {} {\bibfield  {journal} {\bibinfo  {journal}
  {Phys. Rev. Lett.}\ }\textbf {\bibinfo {volume} {82}},\ \bibinfo {pages}
  {4960--4963} (\bibinfo {year} {1999})},\ \bibinfo {note} {[Erratum Phys. Rev.
  {\bf 86}, 4193 (2001)]}\BibitemShut {NoStop}%
\bibitem [{\citenamefont {deVries}()}]{dV2002}%
  \BibitemOpen
  \bibfield  {author} {\bibinfo {author} {\bibfnamefont {J.~C.}\ \bibnamefont
  {deVries}},\ }\href@noop {} {}\bibinfo {howpublished} {Ph.D.~thesis,
  Massachusetts Institute of Technology, Cambridge, MA (2002), URL
  https://dspace.mit.edu/handle/1721.1/4108}\BibitemShut {NoStop}%
\bibitem [{\citenamefont {Berestetskii}\ \emph {et~al.}(1982)\citenamefont
  {Berestetskii}, \citenamefont {Lifshitz},\ and\ \citenamefont
  {Pitaevskii}}]{BeLiPi1982vol4}%
  \BibitemOpen
  \bibfield  {author} {\bibinfo {author} {\bibfnamefont {V.~B.}\ \bibnamefont
  {Berestetskii}}, \bibinfo {author} {\bibfnamefont {E.~M.}\ \bibnamefont
  {Lifshitz}}, \ and\ \bibinfo {author} {\bibfnamefont {L.~P.}\ \bibnamefont
  {Pitaevskii}},\ }\href@noop {} {\emph {\bibinfo {title} {\relax{Quantum
  Electrodynamics, Volume 4 of the Course on Theoretical Physics}}}},\ \bibinfo
  {edition} {2nd}\ ed.\ (\bibinfo  {publisher} {Pergamon Press},\ \bibinfo
  {address} {Oxford, UK},\ \bibinfo {year} {1982})\BibitemShut {NoStop}%
\bibitem [{\citenamefont {Jentschura}\ \emph
  {et~al.}(2017{\natexlab{b}})\citenamefont {Jentschura}, \citenamefont
  {Adhikari},\ and\ \citenamefont {Debierre}}]{JeAdDe2017suppl}%
  \BibitemOpen
  \bibfield  {author} {\bibinfo {author} {\bibfnamefont {U.~D.}\ \bibnamefont
  {Jentschura}}, \bibinfo {author} {\bibfnamefont {C.~M.}\ \bibnamefont
  {Adhikari}}, \ and\ \bibinfo {author} {\bibfnamefont {V.}~\bibnamefont
  {Debierre}},\ }\href@noop {} {}\bibinfo {howpublished} {{\em Long--Range
  Tails in \vdw{} Interactions of Excited--State Atoms: Mixing Terms, Notes and
  Derivations [Supplementary Material for Physical Review Letters]}} (\bibinfo
  {year} {2017}{\natexlab{b}})\BibitemShut {NoStop}%
\bibitem [{\citenamefont {Bethe}(1947)}]{Be1947}%
  \BibitemOpen
  \bibfield  {author} {\bibinfo {author} {\bibfnamefont {H.~A.}\ \bibnamefont
  {Bethe}},\ }\bibfield  {title} {\enquote {\bibinfo {title} {\relax{The
  Electromagnetic Shift of Energy Levels}},}\ }\href@noop {} {\bibfield
  {journal} {\bibinfo  {journal} {Phys. Rev.}\ }\textbf {\bibinfo {volume}
  {72}},\ \bibinfo {pages} {339--341} (\bibinfo {year} {1947})}\BibitemShut
  {NoStop}%
\bibitem [{\citenamefont {Barbieri}\ and\ \citenamefont
  {Sucher}(1978)}]{BaSu1978}%
  \BibitemOpen
  \bibfield  {author} {\bibinfo {author} {\bibfnamefont {R.}~\bibnamefont
  {Barbieri}}\ and\ \bibinfo {author} {\bibfnamefont {J.}~\bibnamefont
  {Sucher}},\ }\bibfield  {title} {\enquote {\bibinfo {title} {\relax{General
  Theory of Radiative Corrections to Atomic Decay Rates}},}\ }\href@noop {}
  {\bibfield  {journal} {\bibinfo  {journal} {Nucl. Phys. B}\ }\textbf
  {\bibinfo {volume} {134}},\ \bibinfo {pages} {155--168} (\bibinfo {year}
  {1978})}\BibitemShut {NoStop}%
\bibitem [{\citenamefont {Bethe}\ and\ \citenamefont
  {Salpeter}(1957)}]{BeSa1957}%
  \BibitemOpen
  \bibfield  {author} {\bibinfo {author} {\bibfnamefont {H.~A.}\ \bibnamefont
  {Bethe}}\ and\ \bibinfo {author} {\bibfnamefont {E.~E.}\ \bibnamefont
  {Salpeter}},\ }\href@noop {} {\emph {\bibinfo {title} {\relax{Quantum
  Mechanics of One- and Two-Electron Atoms}}}}\ (\bibinfo  {publisher}
  {Springer},\ \bibinfo {address} {Berlin},\ \bibinfo {year}
  {1957})\BibitemShut {NoStop}%
\bibitem [{\citenamefont {\relax{C. M. Adhikari {\em et
  al.}}}(2017)}]{AdEtAl2017vdWiii}%
  \BibitemOpen
  \bibfield  {author} {\bibinfo {author} {\bibnamefont {\relax{C. M. Adhikari
  {\em et al.}}}},\ }\href@noop {} {}\bibinfo {howpublished} {{\em Long-range
  interactions of hydrogen atoms in excited states.~III. $nS$--$1S$
  interactions for $n \geq 3$}, in preparation} (\bibinfo {year}
  {2017})\BibitemShut {NoStop}%
\bibitem [{\citenamefont {\relax{R. Pohl {\em et al.} [CREMA
  Collaboration]}}(2010)}]{PoEtAl2010CREMA}%
  \BibitemOpen
  \bibfield  {author} {\bibinfo {author} {\bibnamefont {\relax{R. Pohl {\em et
  al.} [CREMA Collaboration]}}},\ }\bibfield  {title} {\enquote {\bibinfo
  {title} {\relax{The size of the proton}},}\ }\href@noop {} {\bibfield
  {journal} {\bibinfo  {journal} {Nature (London)}\ }\textbf {\bibinfo {volume}
  {466}},\ \bibinfo {pages} {213--216} (\bibinfo {year} {2010})}\BibitemShut
  {NoStop}%
\bibitem [{\citenamefont {\relax{R. Pohl {\em et al.} [CREMA
  Collaboration]}}(2016)}]{PoEtAl2016CREMA}%
  \BibitemOpen
  \bibfield  {author} {\bibinfo {author} {\bibnamefont {\relax{R. Pohl {\em et
  al.} [CREMA Collaboration]}}},\ }\bibfield  {title} {\enquote {\bibinfo
  {title} {\relax{Laser spectroscopy of muonic deuterium}},}\ }\href@noop {}
  {\bibfield  {journal} {\bibinfo  {journal} {Science}\ }\textbf {\bibinfo
  {volume} {353}},\ \bibinfo {pages} {669--673} (\bibinfo {year}
  {2016})}\BibitemShut {NoStop}%
\bibitem [{\citenamefont {Mohr}\ \emph {et~al.}(2016)\citenamefont {Mohr},
  \citenamefont {Newell},\ and\ \citenamefont {Taylor}}]{MoNeTa2016}%
  \BibitemOpen
  \bibfield  {author} {\bibinfo {author} {\bibfnamefont {P.~J.}\ \bibnamefont
  {Mohr}}, \bibinfo {author} {\bibfnamefont {D.~B.}\ \bibnamefont {Newell}}, \
  and\ \bibinfo {author} {\bibfnamefont {B.~N.}\ \bibnamefont {Taylor}},\
  }\bibfield  {title} {\enquote {\bibinfo {title} {\relax{CODATA Recommended
  Values of the Fundamental Physical Constants: 2014}},}\ }\href@noop {}
  {\bibfield  {journal} {\bibinfo  {journal} {Rev. Mod. Phys.}\ }\textbf
  {\bibinfo {volume} {88}},\ \bibinfo {pages} {035009} (\bibinfo {year}
  {2016})}\BibitemShut {NoStop}%
\end{thebibliography}
\end{document}